\documentclass[epj]{svjour}
\usepackage{graphicx,amssymb}
\def\MET{/\hspace{-1.5ex}$E_t$}
\begin{document}
\title{Search for SUSY in Gauge Mediated and Anomaly Mediated Supersymmetry
Breaking Models}
\author{Thomas Nunnnemann for the D\O{} Collaboration
}                     % Do not remove
\institute{Ludwig-Maximilians-Universit{\"a}t M{\"u}nchen, Sektion Physik, 
85748 Garching, Germany}
\date{Received: 15.10.2003}
\abstract{
In this note, recent results on the search for Gauge Mediated Supersymmetry
Breaking (GMSB) and Anomaly Mediated Supersymmetry Breaking (AMSB) at the
LEP and Tevatron colliders are summarized. We report on D\O{}'s search for
GMSB in di-photon events with large missing transverse energy and discuss
the sensitivity of similar searches based on future Tevatron integrated
luminosities.
 % end of PACS codes
} %end of abstract
\authorrunning{T.~Nunnemann}
\titlerunning{Search for SUSY in GMSB and AMSB models}
\maketitle
\section{Introduction}
\label{intro}
If Supersymmetry (SUSY) would be an exact symmetry, SUSY particles would have
the same masses as their Standard Model (SM) partners. 
The required breaking of SUSY
is generally
assumed to occur in a {\em hidden} sector at some very high
energy, and to be mediated to the {\em visible} sector at scales 
$\mathcal{O}(1\,\mathrm{TeV})$
by means of a known interaction. In contrast to Super Gravity (SUGRA), where
SUSY is broken at a very high scale and is mediated by gravity, in 
GMSB models, it is
assumed that at scales as low as $\mathcal{O}(10\,\mathrm{TeV})$ 
the hidden sector is coupled
to a messenger sector, which by
itself is coupled to the visible sector through standard SM gauge 
interactions \cite{Giudice}.

Since those gauge interactions are flavour blind, GMSB models do not generate
unwanted Flavour Changing Neutral Currents (FCNC). 
Another attractive feature of GMSB models is their 
predictiveness having a minimal set of only five parameters
in addition to the SM ones: 
the SUSY breaking scale $\Lambda$,
the messenger mass scale $M_{mess}$,
the number of messenger fields $N_{mess}$,
the ratio of the Higgs vacuum expectation value $\tan\beta$,
and the sign of the Higgs mass term $|\mu|$.

GMSB models have a very distinctive phenomenology. The gravitino $\tilde{G}$
is typically light 
%($\stackrel{<}{\sim}1\,\mathrm{keV}$)
($\lesssim 1\,\mathrm{keV}$)
and is the lightest SUSY particle
(LSP). The next-to-lightest SUSY particle (NLSP) is usually either the lightest
neutralino $\tilde{\chi}_{1}^{0}$, decaying into $\gamma\tilde{G}$, 
or the lightest charged slepton (mostly $\tilde{\tau}_1$), decaying into
$l\tilde{G}$.

In Anomaly Mediated SUSY Breaking (AMSB) models \cite{Randall}, the
rescaling of anomalies in the supergravity Lagrangian gives rise to soft mass
parameters at the visible scale. 
AMSB models are defined by only three parameters and the sign of the Higgs
mass term in addition to the SM. 
The LSP can be either the $\tilde{\chi}^0_1$, which is nearly mass degenerate
with the $\tilde{\chi}^\pm_1$, or the $\tilde{\nu}$, or the $\tilde{\tau}$.

This note briefly reviews searches for both GMSB and AMSB signatures
at LEP and reports on
D\O{}'s preliminary search for GMSB in $\gamma\gamma$ events with large
missing transverse energy (\MET).

\section{Searches for GMSB Topologies at LEP}
Depending on the mass of the LSP and its own mass, the NLSP could have any
decay length relative to detector dimensions. Thus, not only the nature
of the NLSP, but also its lifetime is crucial for the wealth of 
topologies which
can be generated by GMSB models.

\subsection{Topologies with  $\tilde{\chi}_{1}^{0}$ as NLSP}
\label{lepgmsb}
\begin{figure}
  \centering
  \includegraphics[width=0.37\textwidth]{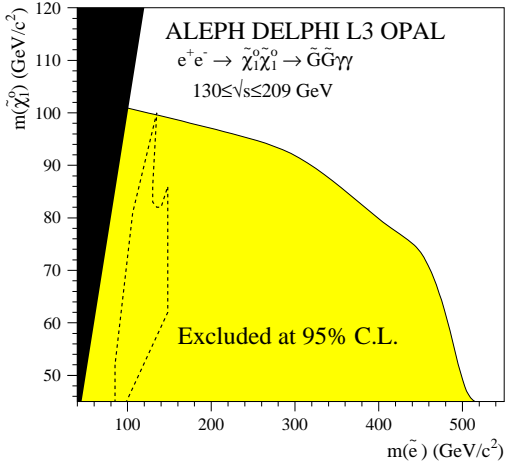}
\caption{95\% exclusion limit for GMSB derived from acoplanar di-photon
analyses at LEP.}
\label{fig:1}       % Give a unique label
\end{figure}

If the first generation neutralino $\tilde{\chi}^0_1$, which can be
pair-produced
at LEP via $\tilde{e}$ exchange,
is short-lived, the event topology will consist
of two acoplanar photons and missing
energy. The main SM background contribution to this channel
is neutrino production with
photon radiation. Since no anomalies in the mass distribution of the
system recoiling against the photons were observed at LEP, limits on the 
neutralino mass $m_{\tilde{\chi}^0_1}$ as function of $m_{\tilde{e}}$ were
set~\cite{lepsusy} and the GMSB interpretation of CDF's $ee\gamma\gamma$\MET \
event~\cite{cdf} was excluded\footnote{All exclusion limits mentioned in this
paper are 95\% C.L.} (see Fig.~\ref{fig:1}).
The LEP collaborations have also searched for $\tilde{\chi}_{1}^{0}$ production
with medium lifetimes by studying topologies with non-pointing photons,
where one of the neutralinos decays within the detector at large distances
away from the interaction point. In the case of long-lived neutralinos, 
searches in
the indirect production channels via slepton-pair or
chargino-pair production have been performed~\cite{opal}.

\subsection{Topologies with $\tilde{\tau}$ or $\tilde{l}$ as NLSP}
\begin{figure}
  \centering
  \includegraphics[width=0.34\textwidth]{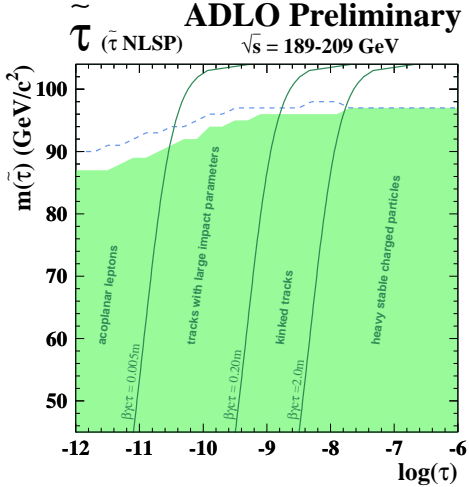}
\caption{Excluded mass as a function of $\tilde{\tau}_1$ lifetime. Expected
limit and lines of equal $\beta\gamma{}c\tau$ are shown, the latter to indicate
the different lifetime regimes.}
\label{fig:2}       
\end{figure}

The LEP collaborations have extensively studied scenarios where the stau
or a slepton is the NLSP. Depending on the NLSP lifetime, different search
strategies have been applied. The combination of those searches covers the
entire range of possible lifetimes~\cite{lepsusy,opal}.
For the $\tilde{\tau}_1$ NLSP scenario, the mass limit as a function of the stau
lifetime is shown in Fig.~\ref{fig:2}, where the sensitivities for the 
various channels are also indicated.
In order of increasing lifetime, searches for the following final states
have been performed:
acoplanar leptons,
tracks with large impact parameter, kinked tracks, and heavy stable
charged particles~\cite{lepsusy,Trigger}.
Combining the results from all four LEP collaborations, a lifetime independent
lower mass limit for the stau mass of $m_{\tilde{\tau}_1}>86.9$\,GeV was
set~\cite{lepsusy}. The combination of searches for all NLSP scenarios was
used by OPAL to perform a complete scan over the parameters of the minimal
model of GMSB~\cite{opal}.

\section{Searches for AMSB Topologies at LEP}
A characteristic feature of AMSB is the small mass difference between
the first-generation
chargino and neutralino, which are mostly gaugino-like.
All LEP experiments have searched for signatures of this model~\cite{lepsusy}.

For mass differences 
$\Delta m = m_{\tilde{\chi}^\pm_1} - m_{\tilde{\chi}^0_1}
\lesssim 200$\,MeV, the lifetime can be so long that the
chargino could be identified through detection of a heavy stable charged
particle or through a kink corresponding to its decay, similar to
the searches for GMSB signatures.
For larger $\Delta m$ those lifetime tags are no longer effective. Since
the visible products of the chargino decay carry little momentum, those
events are both difficult to trigger on and to distinguish them from the large
two-photon background. However, a tag on a photon with large transverse
energy coming from initial state radiation improves both trigger efficiency
and separation from background. Thus, this method achieves the highest
sensitivity in the mass range 
$200\,\mathrm{MeV} \lesssim \Delta m \lesssim 3\,\mathrm{GeV}$.
For even larger values of $\Delta m$ a standard search for chargino decays
can be applied.

Delphi has performed a parameter scan of AMSB models using a combination
of the small $\Delta m$ chargino search together with the search for
$\tilde{\chi}^\pm_1 \rightarrow \tilde{\nu} l^\pm$, the searches for
SM-like or invisible Higgs, and the constraints coming from LEP I measurements
at the $Z$ pole~\cite{Delphi}.

\section{Searches for GMSB Signatures at Tevatron}
The dominating production channels for SUSY particles at Tevatron are 
$\tilde{\chi}^\pm_1 \tilde{\chi}^0_2$ and 
$\tilde{\chi}^+_1 \tilde{\chi}^-_1$ pair production. 

\subsection{Inclusive Search for $\gamma\gamma$\MET}
If the neutralino
is the NLSP and it has a short lifetime, the final state will contain two
photons and large missing transverse energy \MET.

The D\O{} collaboration has done a preliminary search for this GMSB 
signature based on an integrated luminosity of $L=128\,\mathrm{pb}^{-1}$
collected since the start of Run~II. The measurement of the \MET \
spectrum in di-photon events requires two isolated reconstructed photons
with transverse energy $E_T > 20$\,GeV within the acceptance of the
central calorimeter. 
%D\O{}'s highly segmented LAr calorimeter and preshower
%strips allow to point photons to their production vertex.
Two different types of background contributions, 
with and without true \MET,
have been estimated.
The background without true \MET \ (i.e. where the measured \MET \ is due to
resolution effects) are dominantly QCD processes like di-photon or jet
production, where the latter are mis-identified as photons 
(due to leading $\pi^0$'s within the jets). Their contribution is estimated
using a fake $\gamma\gamma$ sample, where at least one photon candidate
fails the shower shape requirement.

The background with true \MET \ (from $\nu$'s) originates from $W$ production
accompanied by a photon or a jet, where the electron from the $W$ decay and
the potential jet are 
misidentified as photons. This contribution is estimated using an 
electron-photon sample and the $e\rightarrow\gamma$ mis-identification
probability derived from data.

Signal acceptance and selection efficiencies have been estimated for various
parameter sets of GMSB
using a full simulation of the D\O{} detector and the photon identification
efficiency, which was determined from data.

\begin{figure}
  \centering
  \includegraphics[width=0.50\textwidth]{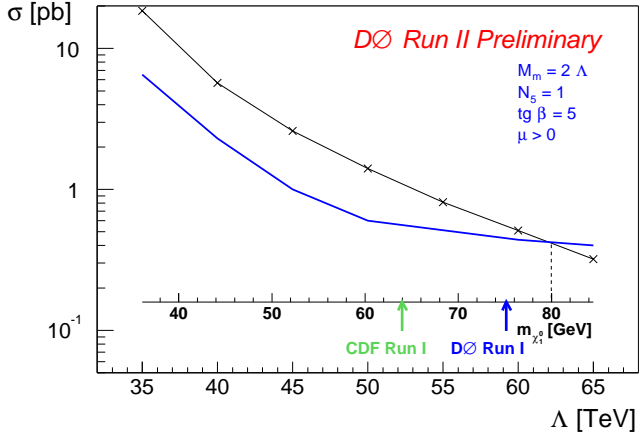}
\caption{Upper cross section limit compared to the 
GMSB prediction for the sum of 
$\tilde{\chi}^\pm_1 \tilde{\chi}^0_2$ and 
$\tilde{\chi}^+_1 \tilde{\chi}^-_1$ pair production and derived lower
mass bound. For comparison the limits obtained at Run~I are indicated.}
\label{fig:3}       
\end{figure}

Since no excess in the \MET \ distribution has been observed, upper
limits on cross sections are calculated based on expected and observed
event rates with \MET \ $>35$\,GeV using a Bayesian approach. By comparing
the derived upper limits for the cross sections with their
theoretical predictions
(shown in Fig.~\ref{fig:3}), a lower limit on the
SUSY breaking scale $\Lambda > 62.5$\,TeV has been derived 
within the parameter set indicated in the figure.
This limit corresponds
to lower bounds on the neutralino and chargino
mass of $m_{\tilde{\chi}^0_1} > 80$\,GeV and 
$m_{\tilde{\chi}^\pm_1} > 144$\,GeV, respectively.

Earlier exclusions interpreted as lower limits on the neutralino mass within
this parameter set are 65, 75, and 100\,GeV from CDF~\cite{cdf}, 
D\O{}~\cite{D0}, and combined LEP (see Sect.~\ref{lepgmsb}) respectively.

A similar preliminary study has been performed by CDF~\cite{Heinemann}.

\subsection{Prospects for Tevatron Run~II}
The discovery reach of the $\gamma\gamma$\MET \ channel has been studied in
Ref.~\cite{Qian} for a similar set of GMSB parameters.  
With an integrated luminosity of $L = 2\,\mathrm{fb}^{-1}$,
GMSB with prompt $\tilde{\chi}^0_1$ decays can be discovered with a C.L. of
$5\sigma$ up to neutralino masses
$m_{\tilde{\chi}^0_1} \approx 165$\,GeV, in excess of the current LEP
exclusion limit.
For intermediate neutralino lifetimes the sensitivity drops as the 
$\tilde{\chi}^0_1$ decays outside the detector, but it can be partly recovered
by including final states with one photon, jets, and missing transverse energy.

\begin{figure}
  \centering
  \includegraphics[width=0.36\textwidth,bb=0 80 384 480,clip=true]{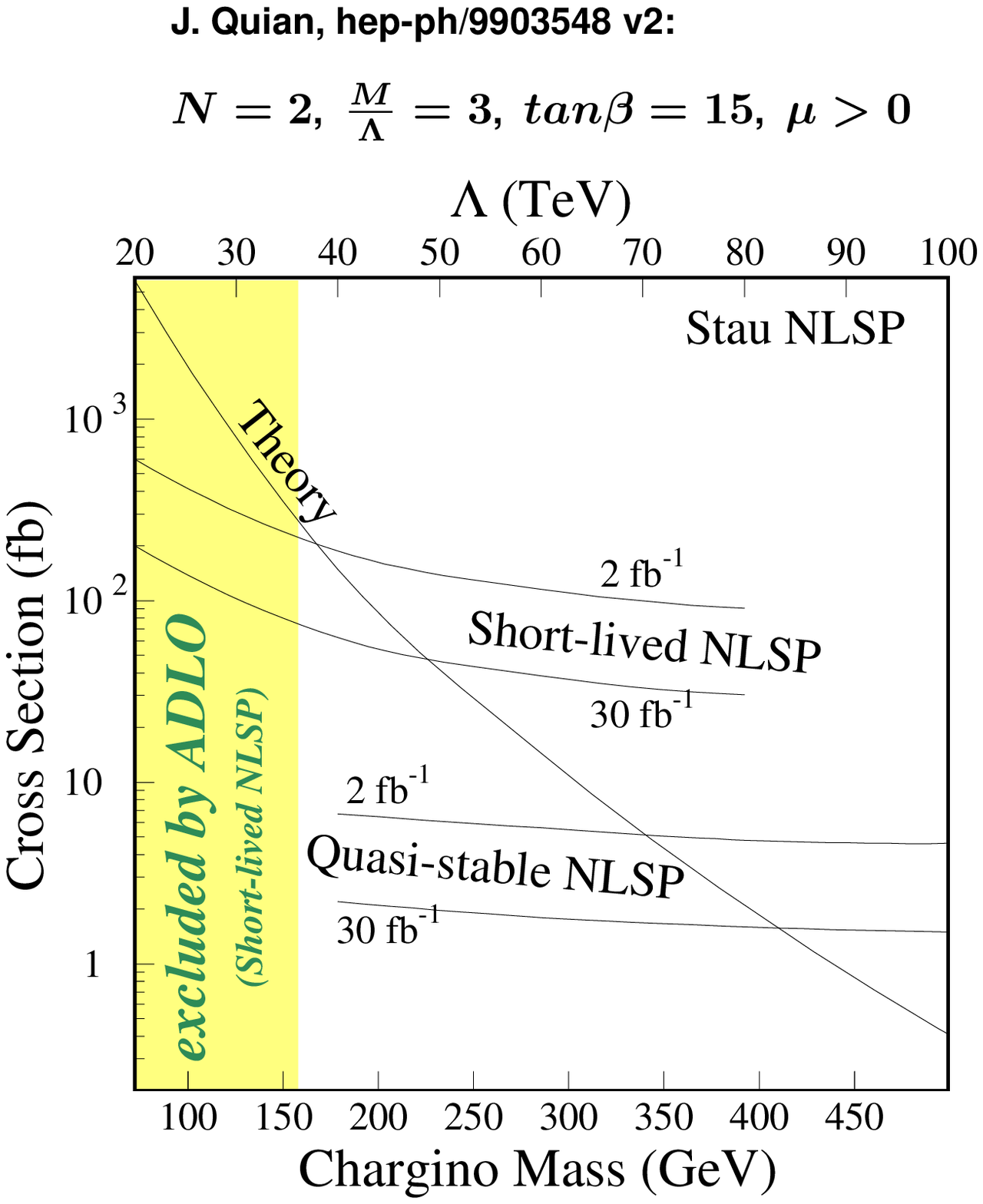}
\caption{LEP exclusion limits for chargino masses compared to prospects
for Tevatron Run~II.}
\label{fig:4}       
\end{figure}

Also for the $\tilde{\tau}$ NLSP scenario, Tevatron has a discovery reach
beyond the current LEP limits as shown in Fig.~\ref{fig:4}~\cite{Qian}.
In case of a
short-lived NLSP the analysis
follows a standard search for a tri-lepton or like-sign di-lepton signature.
If the NLSP is quasi-stable, so that it escapes the detector,
the two staus in the final state can be
identified as two muon-like objects with large ionization energies 
$\mathrm{d}E / \mathrm{d} x$.

\section{Conclusions}
No evidence for the production of SUSY particles 
have been observed at LEP
or Tevatron. At LEP many different topologies as predicted by GMSB and AMSB
have been studied.
The combination of many search channels is used to set limits for all
NLSP lifetimes and to cover most of the kinematically accessible parameter
space for both GMSB and AMSB models. D\O{}'s preliminary
results from Run~II are 
already superseding the GMSB limits obtained from Run~I.
For certain regions of the GMSB parameter region, the experiments at
Tevatron have the potential to significantly improve lower limits
on SUSY particle masses.

\section*{Acknowledgements}
I would like to thank my colleagues from D\O{} and the LEP collaborations for
providing their excellent results.

%\begin{figure*}
%\vspace*{5cm}       % Give the correct figure height in cm
%\label{fig:2}       % Give a unique lab% el
% \end{figure*}
% %
% % For tables use
% \begin{table}
% \caption{Please write your table caption here}
% \label{tab:1}       % Give a unique label
% % For LaTeX tables use
% \begin{tabular}{lll}
% \hline\noalign{\smallskip}
% first & second & third  \\
% \noalign{\smallskip}\hline\noalign{\smallskip}
% number & number & number \\
% number & number & number \\
% \noalign{\smallskip}\hline
% \end{tabular}
% % Or use
% \vspace*{5cm}  % with the correct table height
% \end{table}
%
% BibTeX users please use
% \bibliographystyle{}
% \bibliography{}

\end{document}